\documentclass[review]{elsarticle}

\usepackage{lineno,hyperref,bm,amsmath,epstopdf}
\modulolinenumbers[5]

\begin{document}

\begin{frontmatter}

\title{Nuclear spin-lattice relaxation rate in noncentrosymmetric superconductor Y$_2$C$_3$}


\author[mymainaddress]{Chongju Chen}

\author[mymainaddress]{Biao Jin\corref{mycorrespondingauthor}}
\cortext[mycorrespondingauthor]{Corresponding author}
\ead{biaojin@ucas.ac.cn}

\address[mymainaddress]{School of Physical Sciences, University of Chinese Academy of Sciences - BeiJing 100049 China}

\begin{abstract}
For a noncentrosymmetric superconductor such as Y$_2$C$_3$, we consider a  parity-mixing model composed of spin-singlet $s$-wave and spin-triplet $f$-wave pairing components.
 The $d$-vector in $f$-wave state is chosen to be parallel to the Dresselhaus asymmetric spin-orbit coupling vector.
  It is found that, the quasiparticle excitation spectrum exhibits distinct nodal structure as a consequence of parity-mixing. Our calculation predict anomalous noninteger power laws for low-temperature nuclear spin-lattice relaxation rate $T_{1}^{-1}$.
   We demonstrate particularly that such a model can qualitatively account for the existing experimental results of the temperature dependence of $T_{1}^{-1}$ in Y$_2$C$_3$.
\end{abstract}


\end{frontmatter}

%\linenumbers

\section{Introduction}

\label{intro} The physics of unconventional superconductivity in
materials without inversion symmetry has become a subject of growing
interest \cite{Bauer2012,Yip2014} since the noncentrosymmetric (NCS)
heavy Fermion superconductor CePt$_3$Si was found in 2004
\cite{Bauer2004}. In these materials the superconducting phase
develops in a low-symmetry environment with a missing inversion
center. This broken symmetry generates an antisymmetric spin-orbit
(SO) coupling and prevent the usual even/odd classification of
Cooper pairs according to orbital parity, allowing a mixed-parity
superconducting state \cite{Samokhin2004,Anderson1984}. This mixture
of the pairing channels with different {parities} may result in
unusual temperature and field dependence of experimentally observed
superconducting properties \cite{Bauer2012,Yip2014}.

 For CePt$_3$Si, in particular, where the Rashba-type
\cite{Rashba1960} SO coupling vector  $\bm{\gamma_{k}}\propto(
\hat{k}_{y},-\hat{k}_{x}, 0)$  is generated, various low-energy
thermodynamical and transport properties have been extensively
investigated from both the experimental and theoretical sides. The
NMR relaxation rate \cite{Yogi2004} $T_{1}^{-1}$, thermal
conductivity \cite{Izawa2005}, and London penetration depth
\cite{Bonalde2005} indicate  power law behavior at lowest
temperatures, suggesting the presence of nodal lines in the
quasiparticle excitation spectrum. Besides, the upper critical
magnetic field $H_{c2}$ is  surprisingly large
\cite{Bauer2004,Clogston1962}, and no change in the Knight shift
across the transition temperature $T_{c}$ \cite{Yogi2006} has been
observed. These characteristics are attributed to a spin triplet
superconducting order parameter. Theoretically, Frigeri \textit{et
al.} have proposed an ($s$+$p$)-wave model \cite{Frigeri2004} where
the $d$-vector of $p$-wave state is chosen to be parallel to the SO
coupling vector ($\bm{d_{k}}\propto\bm{g_{k}}$).  The gap function
of this ($s$+$p$)-wave model has the natural form for a system
without inversion symmetry, and  exhibits line nodes when the
$p$-wave pair potential is larger than that of $s$-wave one. It should be
noted that a nonzero $s$-wave pair potential is necessary to get
expected line nodes. Hayashi \textit{et al.} \cite{Hayashi2006} have
demonstrated that the presence of line nodes in this ($s+p$)-wave
model may account for the experimentally observed low-temperature
features of the nuclear spin-lattice relaxation rate $T_{1}^{-1}$ in
CePt$_3$Si on a qualitative level.

The cubic Pu$_{2}$C$_{3}$-type sesquicarbide compound Y$_2$C$_3$  is a NCS superconductor known for its relatively high superconducting transition temperature  \cite{Amano2004} ($T_{c}\sim$18K ). Different from the CePt$_3$Si case,  the Dresselhaus \cite{Dresselhaus1955} SO coupling vector  $\bm{\gamma_{k}}\propto ( \hat{k}_{x} (\hat{k}_{y}^{2}-\hat{k}_{z}^{2}), \hat{k}_{y} (\hat{k}_{z}^{2}
-\hat{k}_{x}^{2}), \hat{k}_{z} (\hat{k}_{x}^{2}
-\hat{k}_{y}^{2}))$ is relevant to Y$_2$C$_3$.

Even many years after its discovery, the nature and symmetry of the superconducting gap function in Y$_2$C$_3$ appears to be full of contradiction. While the specific heat measurement \cite{Akutagawa2007} and tunneling experiment \cite{Ekino2013} are interpreted as a fully gapped isotropic $s$-wave state,  the nuclear spin-lattice relaxation rate \cite{Harada2007} $T_{1}^{-1}$  and muon spin rotation \cite{Kuroiwa2008} ($\mu$SR)  measurements on Y$_2$C$_3$ are qualitatively fitted with a  nodeless two-gap model similar to MgB$_{2}$. On the other hand, Chen \textit{et al.} \cite{Chen2011} have  measured the magnetic penetration depth as a function of temperature and found a weak linear dependence at very low temperatures.  They also reanalysed the NMR data reported in Ref. \cite{Harada2007} and claimed that, where $T_{1}^{-1}\sim T^{3}$ at $T<3$K, as a matter of fact. Such behavior seems to support  the existence of line nodes rather than a fully opened gap in the superconducting state of Y$_2$C$_3$. In addition, the upper critical magnetic field $H_{c2}$ is found to be compatible with the paramagnetic limiting field \cite{Clogston1962,Chen2011}, and {the} Knight shift in NMR \cite{Harada2007} is decreased to approximately  2/3 of its normal-state value. These features are  again incompatible with the single gap or two-gap $s$-wave pictures.
It is expected that  line nodes (or point nodes of second-order) would be generated due to parity-mixing, similar to the case of CePt$_3$Si mentioned above. In order to shed light on these controversy, further   experimental and theoretical studies on the  superconducting properties of Y$_2$C$_3$ are required.

In this work, we  theoretically investigate the nuclear spin-lattice relaxation rate \cite{Harada2007} $T_{1}^{-1}$ on the basis of  ($s$+$f$)-wave model, where the $d$-vector in $f$-wave state is chosen  to be parallel to the Dresselhaus-type asymmetric SO coupling vector. We analyse various possible nodal structures which can be generated by the effect of parity-mixing. In particular, the temperature dependence of the nuclear spin-lattice relaxation rate $T_{1}^{-1}$ is calculated and compared with the experimental  result obtained in Ref. \cite{Harada2007} for Y$_2$C$_3$.
\section{Model Hamiltonian}
Our starting point is the following  mean-field $(s+f)$-wave pairing Hamiltonian

\begin{equation}\label{H}
H=H_{0}+H_{int}.
\end{equation}
The Hamiltonian $H_{0}$  describes the noninteracting conduction electrons in a NCS crystal,

\begin{equation}
H_{0}=\sum_{\bm{k}}\sum_{\alpha,\beta}(\epsilon_{\bm{k}}\sigma_{0}+ \gamma_0\bm{\gamma}_{\bm{k}}\cdot\bm{\sigma})_{\alpha\beta}c^\dagger_{\bm{k}\alpha}c_{\bm{k}\beta},\label{H0}
\end{equation}
where $c^\dagger_{\bm{k}\alpha} (c_{\bm{k}\alpha})$ creates (annihilates) an electron with wave vector $\bm{k}$ and spin $\alpha$, $\bm{\sigma}=(\sigma_{x}, \sigma_{y}, \sigma_{z})$ denotes the vector of Pauli matrices,  $\sigma_{0}$ is the $2\times2$ unit matrix, $\epsilon_{\bm{k}}$  is the parabolic bare band dispersion measured relative to the chemical potential  restricted to $\mid\epsilon_{\bm{k}}\mid<\omega_{c}$, with $\omega_{c}$ being the usual cutoff energy. Furthermore, $\bm{\gamma_{k}}=(\hat{k}_x(\hat{k}_z^2-\hat{k}_y^2), \hat{k}_y(\hat{k}_z^2-\hat{k}_x^2), \hat{k}_z(\hat{k}_x^2-\hat{k}_y^2))$, with $\hat{k}_{x}=\sin\theta_{\bm{k}}\cos\phi_{\bm{k}}$, $\hat{k}_{y}=\sin\theta_{\bm{k}}\sin\phi_{\bm{k}}$,
and $\hat{k}_{z}=\cos\theta_{\bm{k}}$,  is the asymmetric ($\bm{\gamma}_{\bm{k}}=-\bm{\gamma}_{\bm{-k}}$) Dresselhaus SO coupling
vector considered to be relevant for Y$_2$C$_3$ and La$_2$C$_3$. The strength of SO coupling is denoted by $\gamma_{0}$.

The second term  in  Eq. (\ref{H}) represents the pairing interaction:

\begin{eqnarray}
H_{int}=\frac{1}{2}\sum_{\bm{k}}\sum_{\alpha,\beta}[\Delta_{\bm{k},\alpha\beta}c^\dagger_{\bm{k}\alpha}c^\dagger_{\bm{-k}\beta}
+\Delta^\dagger_{\bm{k},\alpha\beta}c_{\bm{-k}\alpha}c_{\bm{k}\beta}\nonumber+\Delta_{\bm{k},\alpha\beta}F^\dagger_{\bm{k},\beta\alpha}], \label{Hint}
\end{eqnarray}
with the anomalous averages $F_{\bm{k},\alpha\beta}=\langle c_{\bm{k}\alpha}c_{\bm{-k}\beta}\rangle$, and the gap function defined by \cite{Mineev1999}

\begin{equation}
\Delta_{\bm{k},\alpha\beta}=-\sum_{\bm{k^{'}}}\sum_{\lambda,\mu}V_{\beta\alpha,\lambda\mu}(\bm{k},\bm{k^{'}})F_{\bm{k^{'}},\lambda\mu},\label{geq}
\end{equation}
where $V_{\alpha\beta,\lambda\mu}(\bm{k},\bm{k^{'}})$ is the pairing potential. In this work, we will adopt  $V_{\alpha\beta,\lambda\mu}(\bm{k},\bm{k^{'}})$ {as} the phenomenological one \cite{Frigeri2006}:

\begin{flalign}
    V_{\alpha\beta,\lambda\mu}(\bm{k},\bm{k^{'}})
    =\nonumber
    -\frac{V_{s}}{2}(i\sigma_{y})_{\alpha\beta}(i\sigma_{y})^{\dagger}_{\lambda\mu}
    -\frac{V_{f}}{2}(\bm{\gamma_{k}}\cdot\bm{\sigma}i\sigma_{y})_{\alpha\beta}(\bm{\gamma_{k^{'}}}\cdot\bm{\sigma}i\sigma_{y})^{\dagger}_{\lambda\mu}\nonumber\\
    +\frac{V_{m}}{2}[ (\bm{\gamma_{k}}\cdot\bm{\sigma}i\sigma_{y})_{\alpha\beta}(i\sigma_{y})^{\dagger}_{\lambda\mu}
    +(i\sigma_{y})_{\alpha\beta}(\bm{\gamma_{k^{'}}}\cdot\bm{\sigma}i\sigma_{y})^{\dagger}_{\lambda\mu}],
\end{flalign}
where the first two terms represent the interaction in the $s$-wave pairing channel and in the spin-triplet $f$-wave pairing channel, respectively,
and the last term describes the scattering between the two channels. In the following, we will chose the interaction parameters $V_{s}$, $V_{f}$, and $V_{m}$ to be positive, and  take for simplicity $V_{m}=\sqrt{V_{s}V_{f}}$ which yields $\Delta_s(T)/\Delta_f(T)=$const. \cite{Frigeri2006}.

Owing to the lack of inversion symmetry, the superconducting gap function Eq. (\ref{geq}) generally contains an admixture of even-parity spin-singlet and odd-parity spin-triplet pairing states,

\begin{equation}
\Delta_{\bm{k},\alpha\beta}=[\psi_{\bm{k}}i\sigma_{y}+\bm{d_{k}}\cdot\bm{\sigma}i\sigma_{y}]_{\alpha\beta}\label{gap},
\end{equation}
where  $\psi_{\bm{k}}=\psi_{-\bm{k}}$ and $\bm{d_{k}}=-\bm{d_{-k}}$ represent the spin-singlet and spin-triplet components, respectively. The direction of the $\bm{d_{k}}$ (the $d$-vector) is assumed to be parallel to $\bm{\gamma_{k}}$, as for this choice the antisymmetric SO interaction is not destructive for spin-triplet pairing\cite{Frigeri2004}.
Hence, we parametrize the $\bm{d}$-vector as $\bm{d_{k}}=\Delta_{f}\bm{\gamma_{k}}$. For the spin-singlet component we assume $s$-wave pairing $\psi_{\bm{k}}=\Delta_{s}$, and choose the amplitudes $\Delta_{s}$ and $\Delta_{f}$ to be real and positive.

Using the vector operator $\Psi_{\bm{k}}=(c_{\bm{k}\uparrow},c_{ \bm{k}\downarrow},c^\dagger_{\bm{-k}\uparrow},c^\dagger_{\bm{-k}\downarrow})^t$, where $(\cdots)^t$ stands for the transposing operation, we can write the Hamiltonian  in a more compact form:

\begin{align}
    H&=\frac{1}{2}\sum_{\bm{k}}\Psi^\dagger_{\bm{k}}\check{H}_{\bm{k}} \Psi_{\bm{k}}+\sum_{\bm{k}}\epsilon_{\bm{k}}
    +\frac{1}{2}\sum_{\bm{k}} \sum_{\alpha,\beta} \Delta_{\bm{k},\alpha\beta}F^{\dagger}_{\bm{k},\beta\alpha},\label{hamiltonian}
\end{align}
where

\begin{align} \check{H}_{\bm{k}}=\left( \begin{array}{cc} \hat{M}_{\bm{k}} &  \hat{\Delta}_{\bm{k}}
        \\ \hat{\Delta}^\dagger_{\bm{k}} & -\hat{M}^*_{\bm{-k}}
    \end{array}
    \right),\label{HM}
\end{align}
with

\begin{align}
    \hat{M}_{\bm{k}}=&\epsilon_{\bm{k}}\sigma_0 + \gamma_0\bm{\gamma}_{\bm{k}}\cdot\bm{\sigma}, \nonumber\\
    \hat{\Delta}_{\bm{k}}=&(\Delta_s+\Delta_{f}\bm{\gamma_{k}}\cdot\bm{\sigma})(i\sigma_y).\label{Delta}
\end{align}
\begin{figure}
    \centering
    \resizebox{0.35 \textwidth}{!}{%
        \centering
        \includegraphics{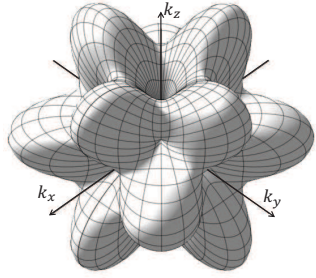}
    }
    %\vspace{5cm}       % Give the correct figure height in cm
    \caption{Schematic illustration of the amplitude of Dresselhaus SO coupling $|\bm{\gamma_k}|$ in $\bm{k}$ space.}
    \label{fig1}       % Give a unique label
\end{figure}
%\begin{figure}[H]
%   \centering
%   \onefigure[scale=0.3]{fig1.eps}
%   \caption{Schematic illustration of the amplitude of Dresselhaus SO coupling $|\bm{\gamma_k}|$ in $\bm{k}$ space.}
%   \label{fig1}
%\end{figure}
\section{Nodal structures}
The Bogoliubov-de Gennes quasiparticle excitation spectrum $E(\bm{k})$ can be obtained readily by diagonalizing the matrix $\check{H}_{\bm{k}}$ above.
One can find four solutions, namely,  $E^{(e)}_{\pm}(\bm{k})$ and  $E^{(h)}_{\pm}(\bm{k})$,  with  $E^{(h)}_{\pm}(\bm{k})=-E^{(e)}_{\pm}(\bm{k})$. We have

\begin{align}
    E^{(e)}_{\pm}(\bm{k})&=\sqrt{(\epsilon_{\bm{k}}\pm\gamma_0|\bm{\gamma_k}|)^2+(\Delta_s\pm \Delta_f |\bm{\gamma_k}|)^2}\equiv E^{\bm{k}}_{\pm}, \label{Ek}
\end{align}
corresponding to two sheets of Fermi surfaces with the energy gaps given by $\Delta_{\bm{k}+}=\Delta_s+\Delta_f|\bm{\gamma_k}|$ and $\Delta_{\bm{k}-}=\Delta_s-\Delta_f|\bm{\gamma_k}|$, respectively. Zeros of $E^{\bm{k}}_{\pm}$ determine the nodal structure of the superconducting state in momentum space. Here let us assume a sufficiently large value of the cutoff energy $\omega_{c}$ $(\omega_{c}\gg\gamma_0, \Delta_s, \Delta_f)$. It is apparent that the upper branch $E^{\bm{k}}_{+}$ is positive definite. Therefore, here we focus on the zeros of the lower branch $E^{\bm{k}}_{-}$.

\begin{figure}
        \includegraphics[scale=0.4]{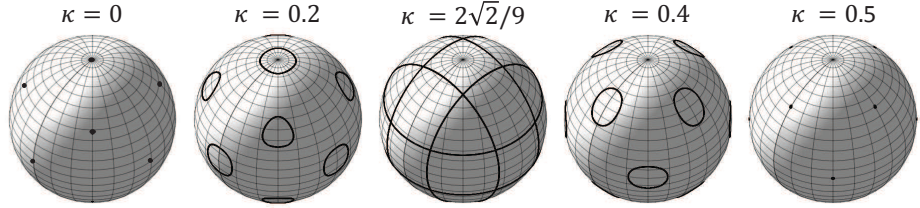}
    \caption{Evolution of nodal structure with the parameter $\kappa$ ($0\leq\kappa\leq0.5$). At $\kappa=0$ ($\kappa=0.5$), $E^{\bm{k}}_{-}$ shows 14 (12) nodal points  of first-order (second-order), while exhibits line nodes for $0<\kappa<0.5$.}
    \label{fig2}       % Give a unique label
\end{figure}

The amplitude of the Dresselhaus SO coupling $|\bm{\gamma_k}|$ (see Fig. \ref{fig1}) becomes zero at 14 points (such as the south and north poles),  possesses 24 saddle points at ($\theta_{k}=\arctan 2\sqrt{2}$, $\phi_{k}=\arccos\sqrt{2}/4$), etc. with $|\bm{\gamma_k}|=2\sqrt{2}/9$, and attains its maximum value 0.5 at 12 points ($\theta_{k}=\pi/2$, $\phi_{k}=\pi/4$), etc. on the Fermi surface.
Therefore, one encounters different nodal topology depending on the ratio $\kappa\equiv\Delta_{s}/\Delta_{f}$. When $\kappa=0$ ($\kappa=0.5$), $E^{\bm{k}}_{-}$ shows 14 (12) nodal points  of first-order (second-order), while exhibits line nodes for $0<\kappa<0.5$ as displayed in Fig. \ref{fig2}. For $\kappa>0.5$, however, we always have $\Delta_{\bm{k}-}\neq0$, and thus the quasiparticle excitation spectrum is gapped.
\section{Nuclear magnetic relaxation rate}
Let us consider the temperature dependence of the nuclear magnetic relaxation rate $T_{1}^{-1}$  defined as
\begin{align}
\frac{1}{T_1T}\propto\sum_{\bm{q}}\frac{\Im[\chi_{-+}(\bm{q},i\omega_n\to \omega+i0^+)]}{\omega}\Bigg|_{\omega\to 0},
\end{align}
where $\Im$ denotes the imaginary part. The dynamical susceptibility in imaginary time is given by
\begin{align}
    \chi_{-+}(\bm{q},i\omega_n)=\int_{0}^{1/T} d\tau\sum_{\bm{k}\bm{k^\prime}}\left\langle \hat{T}c_{\bm{k^\prime-\bm{q}\downarrow}}^{\dagger}(\tau)c_{\bm{k^\prime}\uparrow}(\tau)c_{\bm{k}+\bm{q}\uparrow}^{\dagger}(0)c_{\bm{k}\downarrow}(0)\right\rangle e^{i\omega_n\tau},
\end{align}
 where $\hat{T}$ denotes the time-ordering operator, $\omega_n=(2n+1)\pi T$ is the Matsubara frequency, and $c_{\bm{k}}(\tau)=e^{iH\tau}c_{\bm{k}}e^{-iH\tau}$. We obtain an explicit expression for $T_{1}^{-1}$ as

\begin{flalign}
    \frac{1}{T_1T}\propto\nonumber \sum_{\bm{k,q}}\sum_{\ell,\jmath=\pm}\frac{\delta(E^{\bm{k}}_{\ell}-E^{\bm{q}}_{\jmath})}{T\cosh^2(E^{\bm{k}}_{\ell}/2T)}
    (1+\frac{\epsilon_{\ell,\bm{k}}\epsilon_{\jmath,\bm{q}}+\Delta_{\ell,\bm{k}}\Delta_{\jmath,\bm{q}}}{E^{\bm{k}}_{\ell}E^{\bm{q}}_{\jmath}}),
\end{flalign}
where $\epsilon_{\pm,\bm{k}}=\epsilon_{\bm{k}}\pm\gamma_0 |\bm{\gamma_k}|$. The temperature dependence of $\Delta_{s}$ and $\Delta_{f}$ are
determined by the self-consistent gap equations:
\begin{align}
    \Delta_{s}&=\sum_{\bm{k},\ell}\frac{\tanh( E^{\bm{k}}_{\ell}/2T)}{4E^{\bm{k}}_{\ell}}\Delta_{\ell,\bm{k}}
    (~V_{s}+\ell V_{m}|\bm{\gamma_k}|~), \nonumber\\
    \Delta_{f}&=\sum_{\bm{k},\ell}\frac{\tanh( E^{\bm{k}}_{\ell}/2T)}{4E^{\bm{k}}_{\ell}}
    \Delta_{\ell,\bm{k}}(~V_{m}+\ell V_{f}|\bm{\gamma_k}|~). \label{gap3}
\end{align}
%\begin{figure}[t]
%   \centering
%   \onefigure[scale=0.28]{fig2.eps}
%   \caption{Evolution of nodal structure with the parameter $\kappa$ ($0\leq\kappa\leq0.5$). At $\kappa=0$ ($\kappa=0.5$), $E^{\bm{k}}_{-}$ shows 14 (12) nodal points  of first-order (second-order), while exhibits line nodes for $0<\kappa<0.5$.}
%   \label{fig2}
%\end{figure}
\begin{figure}
    \centering
    \includegraphics[scale=0.5]{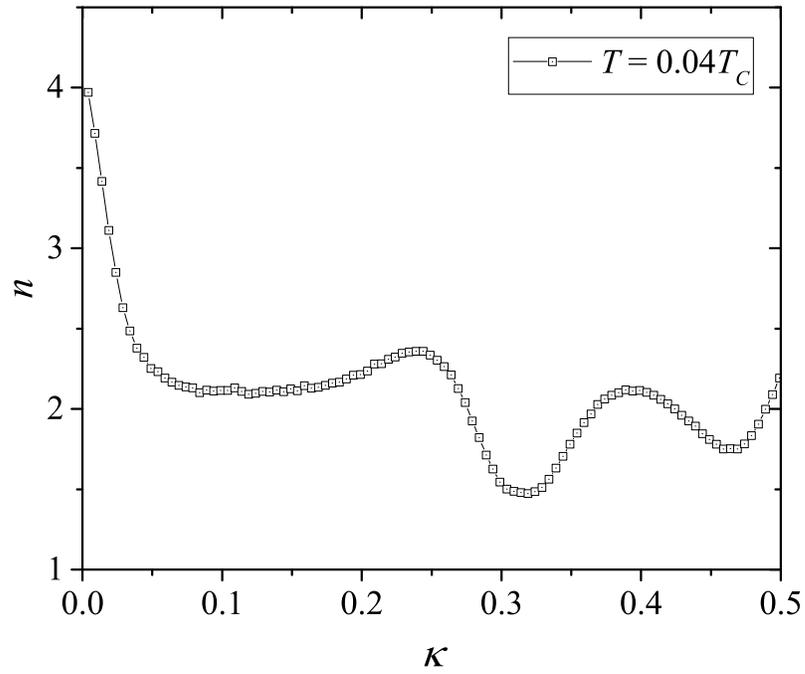}
    \caption{The  exponent of temperature as a function of $\kappa$,  calculated numerically according to $n=d\ln(1/T_1T)/d\ln T$.}
    \label{fignk}
\end{figure}
It turns out that for given values of $V_{s}, V_{f}$, and $\omega_{c}$, $1/T_{1}T$ depends on the $\Delta_{s}$ and $\Delta_{f}$ only through the ratio $\kappa$, and is independent of the strength of SO coupling $\gamma_{0}$, similar to the case of Ref. \cite{Hayashi2006}. It is interesting to see how the low-temperature power law behaviour for  the nuclear spin-lattice relaxation rate $1/T_{1}T\propto T^{n}$ is changed with the ratio $\kappa$.
Plotted in Fig. \ref{fignk} is the exponent of temperature, $n$, as a function of $\kappa$ at $T=0.04T_{c}$ calculated numerically according to $n=d\ln(1/T_1T)/d\ln T$ \cite{Mazidian2013a}. As we see, 
the exponent $n$ attains its maximum  $n=4$ at $\kappa=0$ (point node of first-order), decreases oscillatorily  with increasing $\kappa$, and end with $n\approx2.2$  at $\kappa=0.5$. It is worth noting that, the exponent $n$ is not necessarily to be an integer here, similar to the cases in Ref. \cite{Mazidian2013a}. For $\kappa>0.5$, however, the gap is open and $1/T_{1}T$ decays exponentially in nature.  We present in Fig. \ref{fig3} the temperature dependence of $1/T_{1}T$ obtained experimentally \cite{Harada2007}
by Harada \textit{et al.} for Y$_2$C$_3$, together with the calculated results for $\kappa$=0.47, 0.50, and 0.53 for comparison. Shown in the inset of Fig. \ref{fig3} is the detailed temperate dependence of $\Delta_{s}$ and $\Delta_{f}$ obtained by solving the gap equations Eq. (\ref{gap3}) for $\kappa=0.53$. As can be seen form Fig. \ref{fig3}, there is a fair agreement between our simple theory and experimental results. However, further experimental measurements at low temperatures  $T/T_{c}<0.15$ are needed to obtain a decisive information  about the pairing symmetry and to test the prediction of our theory.
\begin{figure}
    \centering
    \includegraphics[scale=0.5]{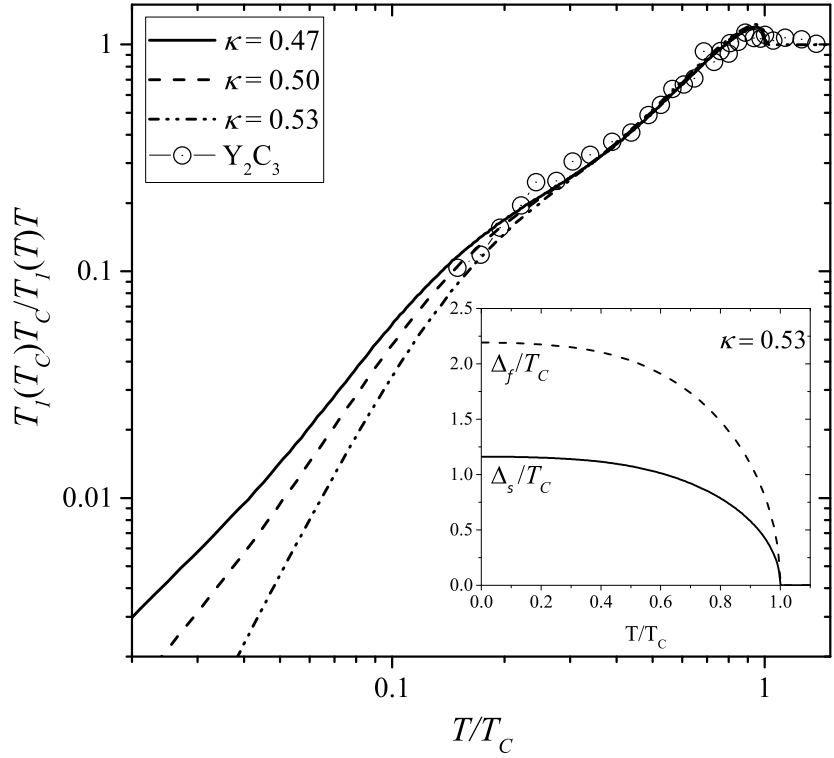}
    \caption{Comparison of the experimental data in Ref. \cite{Harada2007}  with the calculated temperature dependence of $1/T_{1}T$ for $\kappa=$ 0.47, 0.5, and 0.53. Inset shows the temperature dependence of $\Delta_{s}$ and $\Delta_{f}$ at $\kappa=0.53$.}
    \label{fig3}       % Give a unique label
\end{figure}
\section{Summary}
In summary, we have calculated the temperature dependence of the nuclear magnetic relaxation rate $1/T_{1}T$ in  the Dresselhaus-type noncentrosymmetric superconductor Y$_2$C$_3$. We have considered the ($s+f$)-wave parity-mixing model where the $d$-vector is chosen to be parallel to the Dresselhaus SO coupling vector. It is found that various types of nodal structures can be generated due to the effect of parity-mixing, depending on the value of $\kappa$. We also find that, for $\kappa\sim0.5$,  the ($s+f$)-wave model can explain the experimental results fairly well over a wide range of  temperatures. However, accurate measurements of $1/T_{1}T$ at lower temperatures would be crucial to the further clarification of pairing symmetry and gap structure in  Y$_2$C$_3$.

\section{Acknowledgement}
This work is partially supported by Science Research fund of GUCAS (No. Y25102BN00).

%\section*{References}
%\bibliography{Articles_refs}

\end{document}